\documentclass[12pt]{iopart}

\usepackage{amsfonts} 
\usepackage{graphicx} 

\begin{document}

\title[A battery-resistor analogy for further insights on measurement uncertainties]{A battery-resistor analogy for further insights on measurement uncertainties}
\author{Gabriel L. A. de Sousa \&  George C. Cardoso}

\address{Universidade de S{\~a}o Paulo, Department of Physics, FFCLRP, \\
Av. Bandeirantes 3900, Ribeir\~ao Preto, SP, 14040-901, Brazil.}
\ead{gcc@usp.br}
\vspace{10pt}
\begin{indented}
\item[]17 May 2018
\end{indented}

\begin{abstract}

We use analogies to give introductory laboratory students intuition about measurement uncertainties.  Using a battery-resistor circuit we discuss uncertainty concepts and derive  expressions for uncertainty of the mean and sums of uncertainties. Finally, we  draw attention to the fact that the interpretation of standard deviation as uncertainty depends on the statistical distribution of the data, while the interpretation of uncertainty of the mean is largely insensitive to such distribution, especially for large samples. If the resistor in the battery-resistor circuit is a resistive loudspeaker, the uncertainty or noise is literally the acoustic power of the sound produced by such noise. In the sound analogy the statistical distribution of the noise is related to its timbre. 

\end{abstract}

\section{Introduction} 
 Uncertainty analysis is probably one of the least appreciated aspects of the introductory laboratory among students \cite{siegel2007having}. Students need to determine  uncertainty, understand how to add uncertainties  and how repeated measurements or improvements in methodology reduce uncertainty. Despite good books available, e.g.~\cite{hughes2010measurements}, even a plausibility level of understanding the expressions used in experimental uncertainties analysis is normally not accessible without  mathematics that is above the scope of an introductory course.  Giving physical intuition on uncertainties calculations to students would be helpful to increase their level of comfort with the subject \cite{petley1985fundamental}.  Good progress has been made to make measurement and uncertainties more accessible to students \cite{siegel2007having, buffler2008teaching,lubben2001point,buffler2001development,allie2003teaching}; the approaches range from fun activities to nail down the concepts \cite{siegel2007having} to deeper discussions on  ``point" vs. ``set" paradigms \cite{allie2003teaching}. However,  adequate insight into the mathematical formulas used is left for future courses such as statistics and statistical physics, usually without a laboratory context. This scenario often prevents the development of a mental picture about the connection between the measured quantity, the ``set of measurements"  and respective uncertainty analysis and reporting \cite{kung2005teaching}.  Many students are left under the  impression that calculation and reporting of experimental uncertainties is just busywork, not steps towards interpretation of results or further refinements in the experimental system and method.
 %
 
 Measurement is, by definition, the measurement of the mean value \cite{bipm2008evaluation}. In a typical experimental situation we want to \textit{measure} a quantity but only have access to a limited set of readings of the quantity whose true mean (the expected value) we want to estimate. Each given reading in the set most likely differs from the expected value due to random measurement disturbances and to intrinsic fluctuations in the measurement process. Therefore we need to determine both how large this noise is, and the uncertainty it causes in the determination of the mean value as calculated from the finite set of readings.
 
 Traditionally the words uncertainty and error have been used interchangeably. However,  the new guide for uncertainty measurement (GUM) terminology \cite{bipm2008evaluation, kacker2007evolution} recommends the use of the word error only for discrepancies in the expected value due to measurement methodology.  Otherwise, the GUM recommended word is uncertainty. Uncertainties are defined as type A (determined by repeating measurements) or type B (determined from available knowledge about the measurement system and method). This paper focuses, without loss of generality, on  type A uncertainties.  For the sake of simplicity, we will assume that the expected values of experimental errors and type B uncertainties are zero.  
 
 In this article we propose a thought experiment  where we heuristically introduce the notions of mean, standard deviation, uncertainty of the mean and illustrate the need for uncorrelated uncertainties. A simple battery-resistor circuit is used for analogies that help with developing the concepts. Analogies create narratives that stimulate interest and help forming mental models \cite{duit1991role,dagher1995analysis,herrmann1985analogy}. Using the circuit we derive expressions for sums and differences of uncertainties and determine reduction of uncertainty with increased  number of measurements. The electric circuit model also clarifies the need for homoscedasticity (homogeneity of variance) for the derived expressions. Before concluding, we discuss meaning, interpretation and comparisons of quantities with uncertainty. Rudimentary concepts of electric power and summation notation are the only background knowledge needed to follow our arguments. 

\section{Standard deviation is the effective value of noise}
\label{VarAsTheMeanPowerOfNoise}

Measurement uncertainty is a combined characteristic of the quantity intended to be measured, the measurement system and, in many cases, the human operator. In this section we introduce the battery-resistor circuit that will be used for our analogies throughout the paper. Noise, the source of uncertainty, is introduced \textit{ad-hoc} in a thought experiment. We show how uncertainty is related to the noise and how the definitions of variance and standard deviation follow naturally from a measurement problem. 

\begin{figure}[!ht]
\centering
\includegraphics[height=6cm]{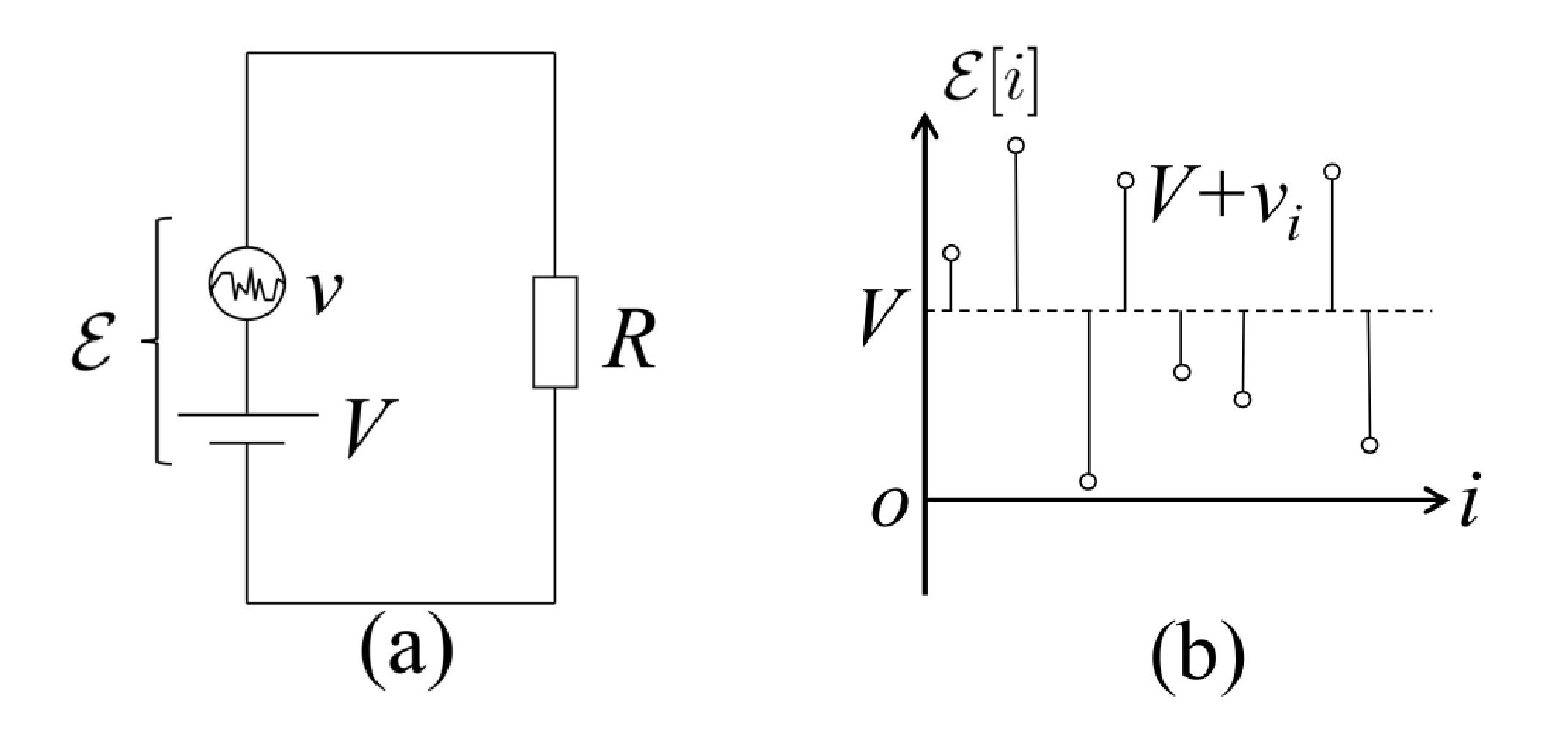}
 \caption{(a) Circuit with battery, noise source and ideal resistive loudspeaker; (b) voltage values  $\mathcal{E}$ on the resistor for  successive measurements, showing the i-th voltage reading. R and V are constants.}
    \label{OneBattery}
\end{figure}

In the circuit of Fig.~\ref{OneBattery}(a), $V$ is an ideal battery and $v$ is an added noise that represents uncertainty in the mean value of the  e.m.f.~$\mathcal{E}$ observed on the resistor $R$. Each time an individual reading $i$ of $\mathcal{E}$ is made,  $v$ assumes a different random voltage, as represented in Fig.~\ref{OneBattery}(b).
The mean value of the e.m.f. on $R$ is $\bar{\mathcal{E}} = V $, where the bar over the variable indicates mean value.   The random noise  $v$ fluctuates around zero and its mean value is assumed to be zero.

 If resistor $R$ is an ideal resistive loudspeaker,  the sound produced by it is  proportional to the measurement noise described above. The constant component $V$ of the battery does not contribute to the loudspeaker noise, since noise is caused only by fluctuations. We will further explore this analogy later.

Let us  calculate the mean power dissipated on the resistor of Fig.~\ref{OneBattery}. The amplitude of the noise $v_i$  changes from  measurement to measurement (or with time) as represented in Fig.~\ref{OneBattery} (b).  From the point of view of resistor $R$, the noise $v$ is the source of uncertainty in the determination of the true mean $\bar{\mathcal{E}}$. Let us assume $R = 1~\Omega$, for simplicity. For a given measurement or instant $i$, the  $i$-th total electric power dissipated on $R$ is given by:

\begin{equation}
    P_i = \frac{\mathcal{E}[i]^2}{R} = \mathcal{E}[i]^2 = (V + v_i)^2,
	\label{eq1}
\end{equation}

\noindent where $\mathcal{E}[i] = V+v_i$ is the total voltage on $R$. Using  Eq.(~\ref{eq1}), the mean electric power over $N$ readings is given by:

\begin{equation} 
    \left\langle  P \right\rangle  =  \frac{1}{N} \sum_{i=1}^N (V + v_i)^2.
    \label{eq2}
\end{equation}

\noindent
Expanding the term $(V + v_i)^2$ in the equation above we get:

\begin{equation} 
    \left\langle  P \right\rangle  =   V^2 + \frac{1}{N} \sum_{i=1}^N v_i^2 + 2V\sum_{i=1}^N \frac{v_i}{N}.
    \label{eq3}
\end{equation}

\noindent
For large $N$ the last term in Eq.(~\ref{eq3}) tends to zero, since the mean value of the noise is zero. We can now distinguish two contributions to the mean power $\left\langle P \right\rangle$: 

\begin{equation} 
    \left\langle  P \right\rangle  = \left\langle P_{signal} \right\rangle + \left\langle P_{noise} \right\rangle.  
    \label{eq4}
\end{equation}

\noindent
The first term is the power $\left\langle P_{signal} \right\rangle$ due to $V$, the ideal battery -- the signal. The second, is $\left\langle P_{noise} \right\rangle$, the contribution of noise $v$. Since $v_i = \mathcal{E}[i] - \bar{\mathcal{E}}$, we can rewrite the term $\left\langle P_{noise} \right\rangle =(1/N) \sum_{i=1}^N v_i^2$ as:

\begin{equation}
    \left\langle P_{noise} \right\rangle= \frac{1}{N}\sum_{i=1}^N (\mathcal{E}[i] - \bar{\mathcal{E}} )^2  \equiv \sigma^2.
    \label{eq5}
\end{equation}

For $N\gg 1$, Eq.~(\ref{eq5}) is identical to the definition of the statistical variance ($\sigma^2$)  of the  e.m.f. $\mathcal{E}$ applied on $R$. Therefore, the variance of the voltage on $R$ can be interpreted as the mean power of the noise. By analogy, the variance $\sigma^2$ of a set of experimental measurements, not necessarily electrical, is the mean power of the fluctuations  around the mean value of the readings. The standard deviation $\sigma$, that has units of the quantity being measured, is the root mean square of the noise power, or effective value of the noise amplitude.

In the analogy above it should have become clear that the standard deviation $\sigma$ is an intrinsic characteristic of the noise. With a larger number of measurements it is possible to better characterize $\sigma$.   The variance is the analog of the acoustic noise power produced by an ideal resistive speaker in place of $R$. 


\section{Sum of uncorrelated uncertainties}
\label{Sum of Uncorrelated uncertainties}

\begin{figure}[!ht] 
    \centering
    \includegraphics[height=7cm]{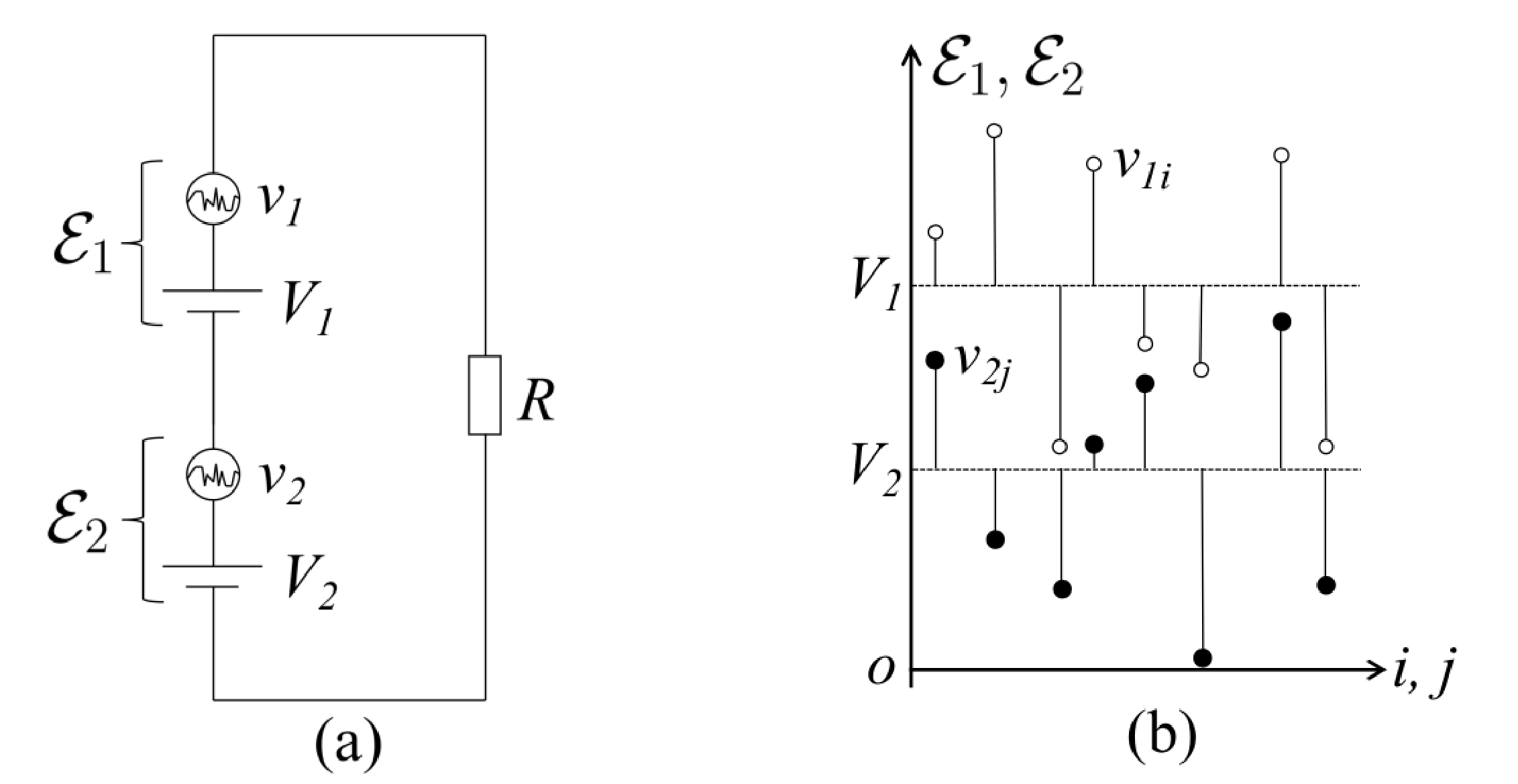}
    \caption{(a) Two voltage sources with noise. (b) Illustration of a series of values of ${\mathcal{E}_1}$ and ${\mathcal{E}_2}$ at measurement events $i$, $j$, where $v_{1i}$ and $v_{2j}$ (not directly measurable) represent fluctuations around the means  $V_1$ and $V_2$.  }
    \label{TwoBattery}
\end{figure}

In this section we will use concepts developed in Section 2 to discuss sums of uncertainties. 
 Figure~\ref{TwoBattery}(a) shows a circuit analogous to the one shown in the previous section but now we have independent ${\mathcal{E}_1}$ and ${\mathcal{E}_2}$  with means $V_1$ and $V_2$ and standard deviations $\sigma_1$ and $\sigma_2$, respectively, to represent a sum of two quantities. 

Figure~\ref{TwoBattery}(b) shows independent readings of ${\mathcal{E}_1}$ and ${\mathcal{E}_2}$ by themselves, displaying the average values $V_1$ and $V_2$ and  noises $v_1$ and $v_2$ for independent events or readings $i$, $j$. Noise sources $v_1$ and $v_2$ could, in principle, be either correlated or uncorrelated. A metaphor for correlated noise vs. uncorrelated noise is a choir vs. a group of students chatting in classroom. The choir has a common conductor and the sounds of its members are  correlated. In the classroom, sounds generated by the students are typically uncorrelated. Below we will show that noises that do not have common causes greatly simplify the determination of sums of uncertainties.

Following the steps of section~\ref{VarAsTheMeanPowerOfNoise}, let us calculate the total power dissipated on $R$ by ${\mathcal{E}_1}$ and ${\mathcal{E}_2}$. For any two readings $i$ and $j$ of the ``noisy" batteries we have that the total power $P_{ij}$ dissipated on  $R = 1~\Omega$ is:

\begin{equation}
    P_{ij} = {\mathcal{E}}[i,j]^2 = (V_1 + V_2 + v_{1i} + v_{2j})^2,
    \label{eq6}
\end{equation}

\noindent where ${\mathcal{E}}[i,j]$ is the total e.m.f.~on R. Indexes $i$ and $j$ are used because $v_1$ and $v_2$ are potentially independent entities. 
The mean total power 
$\left\langle P \right\rangle = \sum_{i,j} P_{ij}/N^2$
dissipated on the resistor is:

\begingroup\makeatletter\def\f@size{10.5}\check@mathfonts
\begin{equation}
    \left\langle P \right\rangle = (V_1 + V_2)^2 + \frac{1}{N^2}\sum_{i=1}^N\sum_{j=1}^N (v_{1i} + v_{2j})^2 + \frac{2(V_1 + V_2)}{N^2} \sum_{i=1}^N\sum_{j=1}^N (v_{1i} + v_{2j}).
    \label{eq7}
\end{equation}
\endgroup

\noindent The last term above tends to zero for large $N$, since $\bar{v_1} =0$ and $\bar{v_2}=0$, similarly to what happened in Eq.~(\ref{eq3}).

Rewriting Eq.~(\ref{eq7}) to put in evidence the variances as defined in Eq.~(\ref{eq5}), we have:

\begingroup\makeatletter\def\f@size{10.5}\check@mathfonts
\begin{equation}
    \left\langle P\right\rangle = (V_1 + V_2)^2 + \frac{1}{N} \sum_{i=1}^N v_{1i}^2 + \frac{1}{N} \sum_{j=1}^N v_{2j}^2 + 2 \left( \frac{1}{N}\sum_{i=1}^N v_{1i} \right) \left( \frac{1}{N}\sum_{j=1}^N v_{2j} \right).
    \label{eq8}
\end{equation}
\endgroup

\noindent 
With uncorrelated uncertainties, the last term with factors in parenthesis tends to zero for large $N$, simplifying the expression. Using definitions of Eq.~(\ref{eq5}), we can rewrite  Eq.~(\ref{eq8}) as:
\begin{equation}
    \left\langle P \right\rangle =
    (V_1 + V_2)^2 +
    \left\langle P_{noise_1} \right\rangle + \left\langle P_{noise_2} \right\rangle.
    \label{eq}
\end{equation}

\noindent
As in section \ref{VarAsTheMeanPowerOfNoise}, the mean power dissipated on $R$ is again a sum of two terms: $\left\langle P_{signal} \right\rangle + \left\langle P_{noise} \right\rangle$, where now $V_1 + V_2$ is the signal. 
 Remembering Eq.~(5), the term $\left\langle P_{noise} \right\rangle$ is the variance $\sigma_T^2$ of the total e.m.f applied to $R$; $\left\langle P_{noise_1}  \right\rangle$ and $\left\langle P_{noise_2}  \right\rangle$ are  the variances $\sigma_1^2$ and  $\sigma_2^2$ of the quantities $\mathcal{E}_1$ and $\mathcal{E}_2$ being added. Therefore the sum of the uncertainties characterized by $\sigma_1$ and $\sigma_2$ is as follows:


\begin{equation}
    \sigma_T^2 = \sigma_1^2 + \sigma_2^2,
    \label{eq9b}
\end{equation}

Observe that while the mean values  of measurements (signals) add up linearly, $\bar{\mathcal{E}} = V_1 + V_2$, the noises or standard deviations add up quadratically: $\sigma_T^2 = \sigma_1^2 + \sigma_2^2$.  Equation~(\ref{eq9b}) is the formula for the sum of uncorrelated uncertainties. 
We would like to draw attention to the fact that in sums, but also in differences $(\mathcal{E}_1 - \mathcal{E}_2)$ between two uncorrelated values  the uncertainties are given by Eq.~(10). To visualize this, just invert one of the batteries in Fig. 2, changing $V_1$ for minus $V_1$ while keeping $V_2$ constant. The total power of the noise is still given by $\sigma_1^2 + \sigma_2^2$, since the powers of the two independent noises on $R$ add irrespective of the polarity of the batteries. These properties will be further explored in the next sections. 

 Before using Eq.~(\ref{eq9b}) it is important to understand its limitations. When the noises $v_1$ and $v_2$ have a common cause, for example, a vibrating air conditioner in the room, the uncertainties will have some degree of correlation. For correlated uncertainties it is easy to show that Eq.~(\ref{eq9b}) is not valid. Let us examine a limiting case where $\sigma_1 = \sigma_2$ but also $v_{1i} = v_{2j}$ (perfect correlation). In this case  we can simply call $v = 2v_1$ in the definition of noise before  Eq.~(\ref{eq5}) and the total uncertainty will be given by $\sigma_T^2 = 4\sigma_1^2$. If the uncertainties were uncorrelated the answer would be $\sigma_T^2 = 2\sigma_1^2$, according to Eq.~(\ref{eq9b}). For partial correlation $\sigma_T^2 = a\sigma_1^2$, with $2\leq a\leq4$.
Correlated uncertainties propagation will not be discussed here and have been treated by Taylor \cite{taylor1985simple}.

\section{Standard deviation of the mean: the uncertainty of the mean value}
\label{StandardError}

In this section we calculate the mean value by averaging  measurements of known standard deviations and  determine the uncertainty of this new mean. The uncertainty in the determination of the mean is called \textit{standard deviation of the mean}, and often incorrectly called \textit{standard error} (SE) \cite{bipm2008evaluation}. In this and in the next sections we will discuss the advantages of use of standard deviation of the mean, as opposed to standard deviation,  to characterize uncertainty in physical sciences.

\begin{figure}[!t] 
\centering
\includegraphics[height=8cm]{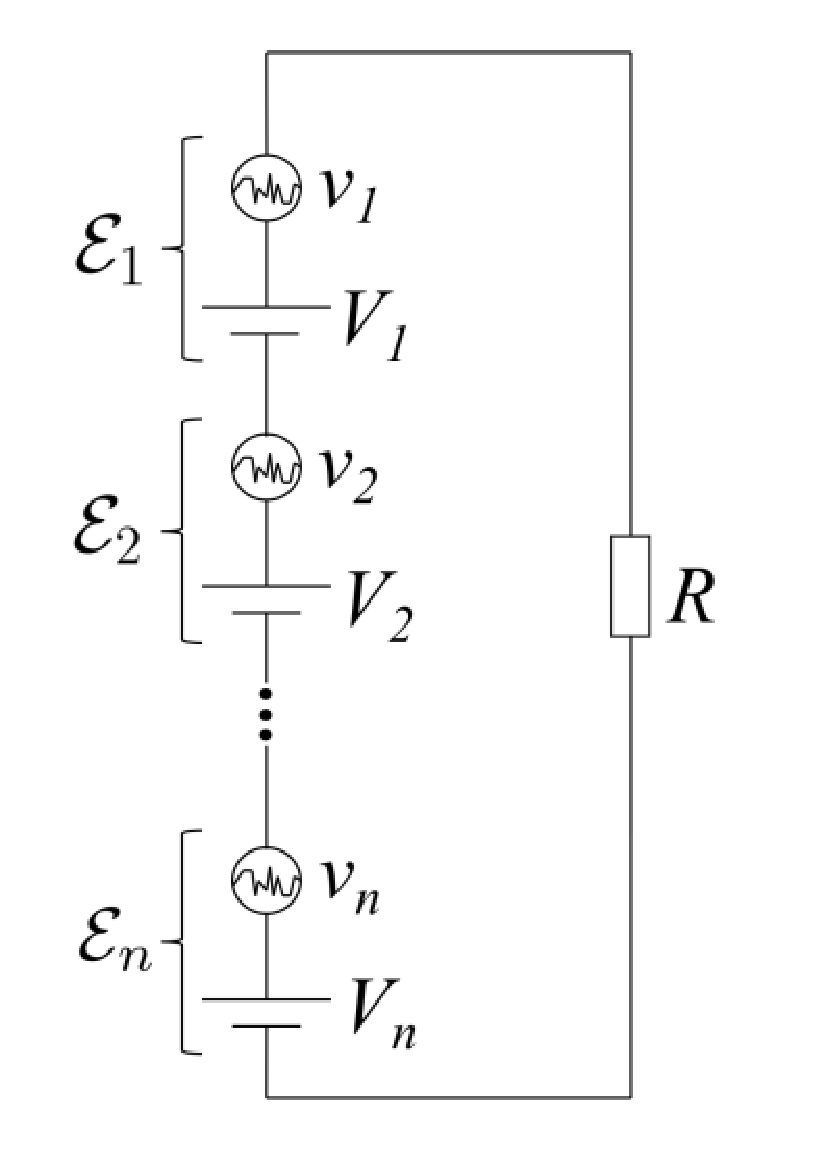} 
\caption{ Sum of values $\mathcal{E}_i$, with mean values $V_i$ and simulated sources of uncertainties $v_i$ (with standard deviations $\sigma_i$). }
\label{MultipleBattery}
\end{figure}

 Figure~\ref{MultipleBattery} is a generalization of Fig.~2(a) for an arbitrary number $n$ of ``noisy" batteries. Following the argument of section \ref{Sum of Uncorrelated uncertainties} we  obtain the following relations for the sum of the means and for the sum of the standard deviations:

\begin{equation}
    \bar{\mathcal{E}} = V_1 + V_2 + V_3 + ... + V_n, 
    \label{eq11}  
\end{equation}

\begin{equation}
    \sigma_T^2 = \sigma_1^2 + \sigma_2^2 + \sigma_3^2 + ... + \sigma_n^2,
    \label{eq12}
\end{equation}

\noindent
where $V_k$  ($k= 1,2,..., n$) are the mean values of voltages of each individual battery, $\bar{\mathcal{E}}$ is the expected mean total voltage on the resistor $R$, and $\sigma_k$ are the respective standard deviations.

Consider now the batteries of Fig.~\ref{MultipleBattery} are all of the same type. The mean voltage of one the batteries of the group will be simply $\bar{V} =\bar{\mathcal{E}}/n$. Since the batteries are of the same type, the individual measurements $V_k$ have identical standard deviations $\sigma$ (same type of noise). We can use Eq.~(\ref{eq12}) and get the uncertainty of the mean  value $\bar{V}$:


\begin{equation}
    \sigma_{\bar{V}} = \frac{\sigma_T}{n} = \frac{\sqrt{n} \, \sigma}{n} = \frac {\sigma}{\sqrt{n}}.
    \label{eq14}
\end{equation}

Equation~(\ref{eq14}) indicates that the uncertainty in the mean value $\bar{V}$, commonly represented by $\sigma_{\bar{V}}$, increases with the standard deviation and decreases with the square root of the number of readings  \cite{reichl1986modern}. 
The \textit{standard deviation} is a characteristic of the noise of  the measurement system, method and/or the intrinsic noise of system being measured. If we want to decrease the \textit{standard deviation of the mean} by increasing the number of measurements, we need, for example,  to increase the number of readings by a factor of a hundred to get an improvement in the uncertainty of the mean by a factor of ten.  If the measurement noise is caused by the measurement system or method, it might be worth investing in improving the experimental conditions to decrease the standard deviation. As seen in Eq.~(\ref{eq14}) a decrease in the standard deviation decreases the uncertainty of the mean much faster than an increase in the number of measurements. 

The number of measurements required to bring uncertainty down to a desired level can be determined after the standard deviation $\sigma$ is estimated  early in the measurement process, in a pilot experiment.
Derivation of Eq.~(\ref{eq14}) assumes that the variances are equal and should not change from measurement to measurement. Therefore to use Eq.~(\ref{eq14}) the multiple measurements of the same quantity should come from the same probability distribution  (homoscedasticity). In the loudspeaker analogy homocedasticity would be equivalent to noises with equal and constant average powers and timbres.


\section{Uncertainty of the mean and statistical distribution of noise}

To measure is to compare. We often need to compare different measurements. When measurement uncertainties are involved, we need to be able to quantify agreement or disagreement between values.  In this section we  discuss the sensitivity of the standard deviation to the statistical distribution of the measurement noise, and the stability of the uncertainty of the mean to the statistical distribution of noise. The statistical distribution of the readings is the timbre of the noise in the loudspeaker analogy. The results of this section are from numerical simulations that we use to illustrate our arguments.

\begin{figure}[!t]
\centering
\includegraphics[width=1\textwidth]{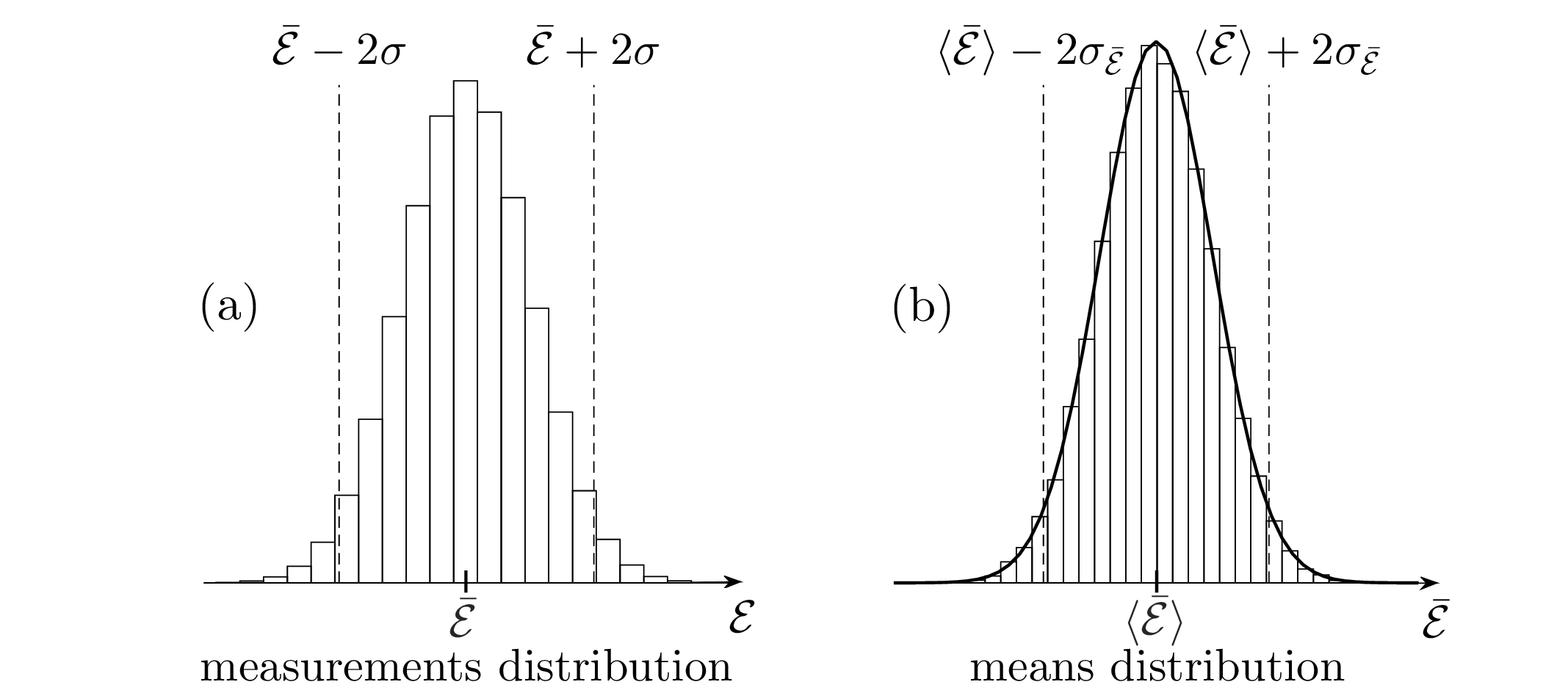}
 \caption{Normal noise. (a) N = 1000 observations. As in any Gaussian distribuition, approximately $95\%$ of the $\mathcal{E}$  values measured are in the interval $\bar{\mathcal{E}}\pm 2\sigma$, between the dashed vertical lines. (b)  n = 2000 repetitions of measurements with  N = 1000 observations each. $\sigma$ is the standard deviation of $\mathcal{E}$, $\left<\bar{\mathcal{E}}\right>\approx \bar{\mathcal{E}}$ and $\sigma_{\bar{\mathcal{E}}} \approx \sigma/\sqrt{2000}$. Once again,  approximately $95\%$ of the $\bar{\mathcal{E}}$ means are found between the two vertical lines. The adjusted line on (b) is a Gaussian fit.}
    \label{gaussiana}
\end{figure}

\begin{figure}[!t]
\centering
\includegraphics[width=1\textwidth]{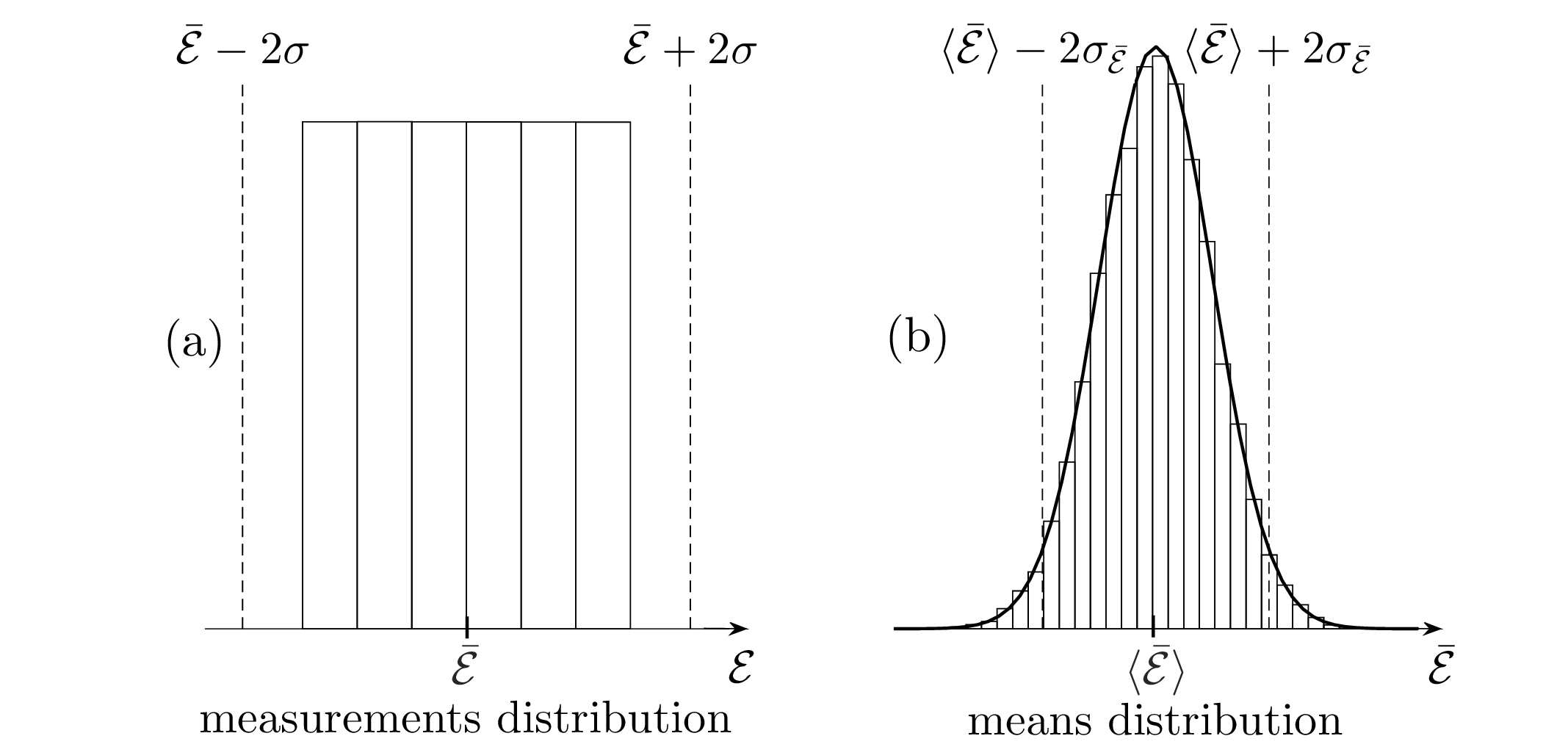}
 \caption{Uniform distribution of six-sided dice.  (a) N = 1000 observations; the six bins are the probabilities of outcomes 1 through 6, with mean $\bar{\mathcal{E}}$ very close to true exact value  3.5 and $\sigma = 1.708$. The dashed lines at $\bar{\mathcal{E}}\pm 2\sigma$ are expected at positions 0.084 and 6.916, outside of the 1-6 range of the uniform distribution. (b)  n = 2000 repetitions of the experiment to determine $\bar{\mathcal{E}} $ with  N = 1000 observations in each repetition. $\sigma$ is the standard deviation of $\mathcal{E}$, $\left<\bar{\mathcal{E}}\right>\approx \bar{\mathcal{E}}$ and $\sigma_{\bar{\mathcal{E}}} \approx \sigma/\sqrt{2000}$. The adjusted line on (b) is a Gaussian fit.}
    \label{uniforme}
\end{figure}

In Fig.~\ref{gaussiana}(a) we show the histogram of measurement values whose noise has the so-called normal or Gaussian statistical  distribution. ${\bar{\mathcal{E}}}$ is the mean value of the set of readings and the dashed lines represent the mean value plus or minus two standard deviations.
The region between the dashed lines represent approximately 95\% of the events in a Gaussian distribution. Therefore, on average, in measurements obeying the normal distribution, only around 5\%  of observations will fall outside the range ${\bar{\mathcal{E}}} \pm 2\sigma$.

Figure~\ref{gaussiana}(b) shows the distribution of the mean values $\bar{\mathcal{E}}$, where each mean value is determined from a distribution such as the one of Fig.~\ref{gaussiana}(a). The width of ~\ref{gaussiana}(b)  is proportional to the standard deviation of the mean. On average, in 95\% of the  experiments with $\mathcal{E}$ noise distribution such as in Fig.~\ref{gaussiana}(a), the mean value obtained will be  in the range $\bar{\mathcal{E}} \pm 2\sigma_{\bar{\mathcal{E}}}$. The mean value of the means distribution $\left<{\bar{\mathcal{E}}}\right>$ is the same as the expected value ${\bar{\mathcal{E}}}$, for very large $N$; actual values might be slightly different due to statistical fluctuations for a finite number of readings.

Figure 5(a) shows a histogram  of readings where the statistics is not Gaussian. The distribution displayed is a uniform distribution of readings of unbiased dice, where the probabilities of integer outcomes 1 through 6 are identical. This truncated type of distribution is common in cases where there are natural physical limitations or when the quantities under consideration are pre-sorted such as in quality control in a factory. The positions of the calculated ${\bar{\mathcal{E}}} \pm 2\sigma$ for the  distribution shown are represented in Fig. 5(a) by the vertical dashed lines.  Observe that, different from the Gaussian distribution, here we are 100\% certain that any measurement will lead to a value $\mathcal{E}$ well inside the ${\bar{\mathcal{E}}} \pm 2\sigma$  interval. If it was a Gaussian distribution, approximately 5\% of the observations would fall outside the ${\bar{\mathcal{E}}} \pm 2\sigma$ interval. This shows that the interpretation of \textit{standard deviations} for measurements depends on the statistical distribution of the noise.
For each experimental case we need to have an estimate of the mathematical form of the noise distribution before interpreting the meaning of the standard deviations in  statements such as $\bar{x} \pm \sigma$.

Figure 5(b) shows a distribution of mean values when each  ${\bar{\mathcal{E}}}$  comes from an experiment with noise distribution as shown in Fig. 5(a).  Observe that the distribution in 5(b) looks Gaussian, despite the fact that 5(a) is a uniform distribution. The central limit theorem \cite{reichl1986modern} explains what is happening: if a random variable is measured repeatedly and independently and results are averaged, the distribution of the averages tends to a normal distribution for a large number of measurements.  Therefore, independently of the statistical distribution of the  noise, the distribution of the means tends to a Gaussian distribution. The interpretation of uncertainty written as \textit{standard deviations of the mean}  follows normal distribution rules for measurements with a large enough number of readings. In practice, with $N$ starting at around 16 the Gaussian formalism may be used for the uncertainty of the mean,  since in this case there is only one  significant figure used for the uncertainty.  The limitation in the number of significant figures in the uncertainty of the mean arises because the estimation of the standard deviation also suffers from uncertainty. There is an uncertainty of approximately $71/\sqrt{(N-1)}$ \% in the determination of $\sigma_{\bar{\mathcal{E}}}$. For example, if N = 50, there is approximately 10\% uncertainty in the determination of $\sigma_{\bar{\mathcal{E}}}$ itself; more than one or two figures to represent $\sigma_{\bar{\mathcal{E}}}$ would not be significant or meaningful. For a discussion on how to display the number of significant figures in the uncertainty of the mean, based on the number of measurements, see reference~\cite{hughes2010measurements}.

The advantage of using the uncertainty of the mean instead of the standard deviation of a distribution is to make the uncertainty analysis independent of the probability distribution. The dashed lines in Fig. 5(b) represent the limits ${\bar{\mathcal{E}}} \pm 2\sigma_{\bar{\mathcal{E}}}$. Each time we perform the experiment with a finite number of readings the mean will change slightly.  
 Independently of the statistical distribution of the noise, approximately 95\% of the experiments performed with the same experimental method and same number of measurements and will give \textit{means} in the range ${\bar{\mathcal{E}}} \pm 2\sigma_{\bar{\mathcal{E}}}$. 
 
\begin{figure}[!ht]
\centering
\includegraphics[width=0.8\textwidth]{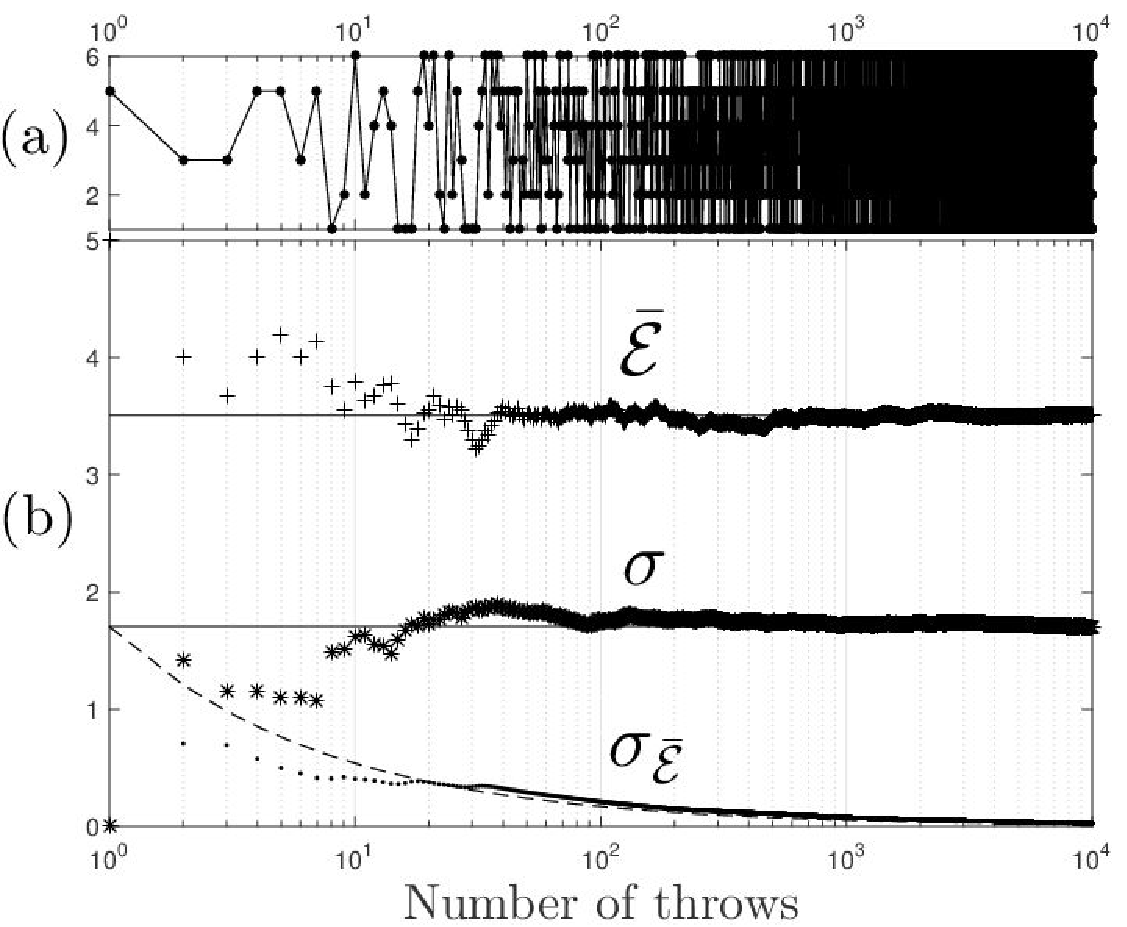}
 \caption{Dice throwing experiment. (a) Values observed in successive throws of dice in an experiment, (b) The crosses are the averages ${\bar{\mathcal{E}}}$ of the values (a) up to the $n$th observation, the stars are the standard deviations  $\sigma$ calculated up to the $n$th observation, and the dots are the standard deviations of the means $\sigma_{\bar{\mathcal{E}}}$, up to the $n$th mean ${\bar{\mathcal{E}}}$. The dashed line is the expected dependency  $\sigma/\sqrt{n}$ for $\sigma_{\bar{\mathcal{E}}}$. Each of the $n$ throws comes from a uniform distribution with  standard deviation $\sigma$.  }
    \label{dice}
\end{figure}
 
 Figure 6 shows a different perspective on the data presented in the distributions of Fig. 5. Figure 6(a) shows a series of experimentally observed values of six-sided dice being thrown. The values $\mathcal{E}$ observed are integers between 1 and 6. The theoretical expected value for the mean value $\bar{\mathcal{E}}$ of the dice results is exactly 3.5. Panel 6(b) shows the results for $\bar{\mathcal{E}}$, $\sigma$ and $\sigma_{\bar{\mathcal{E}}}$ obtained from the ``experimental" values shown in Fig. 6(a), from throw 1 up to the Nth throw. We observe in 6(b) how the mean value fluctuation decreases with the number of measurements (throws) and converges to 3.5 as the number of measurements increases. The standard deviation $\sigma$ behaves in a similar way: the uncertainty in $\sigma$ decreases with the number of measurements and $\sigma$ converges to its expected value of approximately 1.7. 
 Finally the standard deviation of the mean $\sigma_{\bar{\mathcal{E}}}$ also decreases with the number of measurements and tends to zero for a very large number of measurements. Observing the plots of $\mathcal{E}$ and $\sigma_{\bar{\mathcal{E}}}$ we can see a decrease in the uncertainty of the standard deviation of $\bar{\mathcal{E}}$, that is given by $\sigma_{\bar{\mathcal{E}}}$  . The dashed line in the  $\sigma_{\bar{\mathcal{E}}}$ plot is the theoretically expected decrease in the uncertainty of the mean  with the number of experiments, $\sigma_{\bar{\mathcal{E}}} = \sigma/\sqrt{n}$.
 In a different run of the same dice experiment the curves would be slightly different, especially for the low $N$ region, but the trends would be the same.
 
 In the dice experiment the standard deviation is an intrinsic characteristic of the observed data, not of the experimental method; only an increase in the number of measurements can decrease the uncertainty in an experimental determination of the mean value $\bar{\mathcal{E}}$. On the other hand, in experimental cases where the measurement noise is also due to the method used, a decrease in $\sigma$ by improvement of the experimental conditions would lead to a faster convergence of the mean value since $\sigma_{\bar{\mathcal{E}}}$ depends linearly on $\sigma$ but decreases only with the inverse of the square root of the number of measurements. 

\section{Comparing  quantities that have uncertainties}

We have seen in the previous sections that we can interpret \textit{uncertainty of the mean} assuming a Gaussian distribution whenever the number of measurements averaged is large. This is an advantage of use of  uncertainty of the mean compared to standard deviation. Therefore, let us focus here on uncertainties described in terms of uncertainties of the means.

Let us say we have two values $X =  \bar X  \pm 2\sigma_{\bar{X}}$ and $ Y = \bar Y  \pm 2\sigma_{\bar{Y}}$.
Does X agree or disagree with Y?
Due to the probabilistic nature of the measurements, in general we cannot claim with 100\% certainty that two values with uncertainty agree or disagree with each other. However, we can estimate the percentage of the cases where similar experiments, if repeated, will lead to a specified degree of agreement or disagreement with each other.

If $X$ agrees with $Y$, then the difference $X-Y$ must have a very low probability of not being zero.Of course, $X$ and $Y$ agree 100\% if $X-Y$ \textit{is} zero, with no uncertainty -- but that is never the case when uncertainties are present.
In Eq.~(10) and discussion thereafter, we have seen that the difference between two uncertainties is also given by said equation. Therefore, $X$ and $Y$ agree in the confidence interval of 95\% if the difference $(\bar X  - \bar Y) \pm 2\sqrt{\sigma_{\bar{X}}^2 + \sigma_{\bar{Y}}^2}$ contains the value zero. 
In this case only around 5\% of similar experiments with the same number of readings will find $(X - Y)$ outside the interval $(\bar X  - \bar Y) \pm 2\sqrt{\sigma_{\bar{X}}^2 + \sigma_{\bar{Y}}^2}$.  

Let us take as an example two mass quantities:  $X = (100 \pm 4)$ kg, 95\%  Confidence Interval (uncertainty approximately $2\sigma_{\bar{X}}$) and 
 $Y = (106 \pm 3)$ kg, 95\% C.I. The difference between $X$ and $Y$ is
 $(6 \pm 5)$ kg, 95\% C.I.
 Since this difference does not include zero, $X$ and $Y$ are statistically different at the 95\% Confidence Interval. However with a higher confidence interval we might not be able to say that $X$ and $Y$ are statistically different. In general, we can not say with 100\% confidence that two mean values with uncertainties are distinct. Now even if we improve the measurements and get lower uncertainties, we might have, for example $W = (100.0 \pm 0.4)$  kg, 95\% C.I. and $Z = (106.0 \pm 0.3)$ kg, 95\% C.I., and the difference  is $(W - Z) = (6.0 \pm 0.5)$ kg, 95\% C.I. Even if we multiply the uncertainty by 3, making a  $6\sigma$ level uncertainty that  increases the confidence interval to 99.9999998\%, we can still have that $W$ and $Z$ can be considered distinct, since the confidence interval does not include zero: $(6.0 \pm 1.5)$ kg, 99.9999998\%  C.I. At this level of confidence, repetitions of the experiment will have $(W - Z)$ outside the range  $(6.0 \pm 1.5)$ less than once in every 500 million repetitions. In this case we say that $W$ and $Z$ are distinct for all practical purposes. 
 
 Different fields of Science have different requirements on confidence intervals. While applied sciences are less stringent, using 95.45\% and 99.73\% for confidence intervals ($2\sigma$ and $3\sigma$), fundamental physics uses at least $5\sigma$ standard to accept a physical discovery (99.999942\% C.I.). More systematic and finer comparisons between experimental quantities can be done with help of statistical hypothesis testing.

Before closing, we would like to remind the reader that the  statistical uncertainties discussed in this paper do not account for discrepancies (errors) in the measured values due to the methodology used -- type B uncertainty due to measurement equipment reproducibility issues.  To establish whether the measurement method used was adequate, alternative measurement methods need to be used for cross checking. 
On a different note, our analogies are directly valid for sums and differences of quantities with uncertainties. We showed how to determine the uncertainty $\sigma_{\bar{f}}$ in an expression of the type $f(x,y) = x \pm y$, where both $x$ and $y$ have uncertainties. The determination of the influence of $\sigma_{\bar{x}}$ and  $\sigma_{\bar{y}}$ on $\sigma_{\bar{f}}$ for an arbitrary function $f(x,y)$ can be visualized from an extrapolation of our analogies. For uncorrelated uncertainties, we just need to determine how the noise power in $f(x,y)$  depends on $\sigma_{\bar{x}}$ and on $\sigma_{\bar{y}}$, independently, and add these contributions, since the total power of noise at the resistor of the analogy is the independent sum of the partial powers due to each noise source. For small uncertainties, this calculation is usually done using partial derivatives to determine the independent contributions to the noise in $f$ due to each uncertainty $\sigma_{\bar{x}}$ and  $\sigma_{\bar{y}}$, and adding these contributions in quadrature \cite{hughes2010measurements}. Alternatively, the contributions can be determined via finite differences as follows: $\sigma_{\bar{f}}^2 =|f(\bar{x},\bar{y}) - f(\bar{x}\pm\sigma_{\bar{x}},\bar{y})|^2 + |f(\bar{x},\bar{y}) -f(\bar{x},\bar{y}\pm\sigma_{\bar{y}})|^2$, where, for very small uncertainties, either the positive or negative signs can be used . 

\section{Conclusion}
We have proposed a circuit analogy to help introducing basic concepts in measurements and uncertainties. The analogy drew attention to the need for uncorrelated sources of uncertainty and for homoscedesticity in use of typical expressions used in the introductory laboratory. The loudspeaker analogy  clarifies the fact that the standard deviation is a property of the  measurement noise, not of the number of points or readings used in the measurement.  In the battery-resistive loudspeaker metaphor,  the acoustic power of the noise is the variance of quantity being measured;  the statistical distribution of the noise is the analogous of the timbre of the noise in the speaker. Finally we discussed how the interpretation of uncertainty of the mean is much less susceptible to the statistical distribution of the experimental noise compared to standard deviation.

\section{Acknowledgements}
We acknowledge Dr. Gilberto Nakamura for critical reading of the manuscript. G.C.C. acknowledges funding from PAJT/CAPES 88881.067978/2014-01.

\section{References}
\bibliographystyle{vancouver} 
\bibliography{Referencias} 
\end{document}